\documentclass[aps,nofootinbib,prd,preprintnumbers,showpacs]{revtex4}
\usepackage{eurosym}
\usepackage{multirow}
\usepackage{wrapfig}
\usepackage{amsfonts}
\usepackage{amsmath}
\usepackage{hyperref}
\usepackage{graphicx,color}
\usepackage{amssymb}
\usepackage[english]{babel}
\usepackage{graphicx}
\usepackage{epsfig}
\usepackage{bm}
\usepackage{longtable}
\usepackage{verbatim}
\usepackage{longtable}
\usepackage{color}
\usepackage{subfigure}
\usepackage[utf8]{inputenc}

\newcommand{\hs}{\hspace*{0.3cm}}

\newcommand{\ms}{\hspace*{1cm}}
\newcommand{\be}{\begin{equation}}
\newcommand{\ee}{\end{equation}}
\newcommand{\bea}{\begin{eqnarray}}
\newcommand{\eea}{\end{eqnarray}}
\newcommand{\ben}{\begin{enumerate}}
\newcommand{\een}{\end{enumerate}}
\newcommand{\bit}{\begin{itemize}}
\newcommand{\eit}{\end{itemize}}
\newcommand{\bde}{\begin{widetext}}
\newcommand{\ede}{\end{widetext}}
\newcommand{\nn}{\nonumber}
\newcommand{\crn}{\nonumber \\}

\newcommand{\al}{\alpha}
\newcommand{\la}{\lambda}
\newcommand{\bet}{\beta}

\newcommand{\va}{\varphi}
\newcommand{\om}{\omega}

\newcommand{\fr}{\frac}

\newcommand{\bc}{\begin{center}}
\newcommand{\ec}{\end{center}}
\newcommand{\Ga}{\Gamma}
\newcommand{\de}{\delta}
\newcommand{\De}{\Delta}

\newcommand{\varep}{\varepsilon}

\newcommand{\La}{\Lambda}
\newcommand{\si}{\sigma}

\newcommand{\mathsym}[1]{{}}

\input{tcilatex}
\begin{document}

\title{The first $\De(27)$ flavor 3-3-1 model with low scale seesaw mechanism}
\author{A. E. C\'arcamo Hern\'andez${}^{a}$}
\email{antonio.carcamo@usm.cl}
\author{H. N. Long$^{b}$}
\email{hnlong@iop.vast.ac.vn}
\author{V. V. Vien$^{d,e}$}
\email{wvienk16@gmail.com}
\affiliation{$^{{a}}$Universidad T\'{e}cnica Federico Santa Mar\'{\i}a and Centro Cient%
\'{\i}fico-Tecnol\'{o}gico de Valpara\'{\i}so, \\
Casilla 110-V, Valpara\'{\i}so, Chile,\\
$^{{b}}$Institute of Physics, Vietnam Academy of Science and Technology, 10
Dao Tan, Ba Dinh, Hanoi Vietnam\\
$^{{d}}$ Institute of Research and Development, Duy Tan University, 182
Nguyen Van Linh, Da Nang City, Vietnam\\
$^{{e}}$ Department of Physics, Tay Nguyen University, 567 Le Duan, Buon Ma
Thuot, DakLak, Vietnam}
\date{\today }

\begin{abstract}

We propose a viable model based on the $SU(3)_C\times SU(3)_L\times U(1)_X$ gauge group, augmented by the $U(1)_{L_g}$ global lepton number symmetry
and the $\De (27) \times Z_3\times Z_{16}$ discrete group, capable of explaining the Standard Model (SM) fermion masses and mixings, and having a low
scale seesaw mechanism which can be tested at the LHC. In addition the model provides an explanation for the SM fermion masses and mixings.
In the proposed model, small masses for the light active neutrinos are generated by an inverse seesaw mechanism caused by non renormalizable
Yukawa operators and mediated by three very light Majorana neutrinos and the observed hierarchy of the SM fermion masses and mixing angles
 is produced by the spontaneous breaking of the $\De (27) \times Z_3 \times Z_{16}$ symmetry at very large energy scale.
 This neutrino mass generation mechanism is not presented in our previous 3-3-1 models with $\De(27)$ group \cite{Vien:2016tmh,Hernandez:2016eod},
 where the masses of the light active neutrinos arise from a combination of type I and type II seesaw mechanisms \cite{Vien:2016tmh} as well as from
 a double seesaw mechanism  \cite{Hernandez:2016eod}. Thus, this work corresponds to the first implementation of the $\De(27)$ symmetry in a 3-3-1
 model with low scale seesaw mechanism.

\end{abstract}

\maketitle

\section{Introduction}

  Despite its great successes, the SM  still has some puzzles such as  the smallness of neutrino masses, Dark Matter, etc. In addition,
the SM does not render an agreeable explanation for the fermion masses and mixings. It is well known that that the top quark mass (around 175 GeV) is 13
 orders of magnitude much larger than the light active neutrino masses. Furthermore, the mixings among quarks are small while lepton mixings are quite large.
 Moreover, two of the leptonic mixing angles are large and the another one is Cabibbo sized.

In traditional way, the particle masses are generated through Yukawa couplings, and the latter also enter in the CKM/PMNS matrices. Thus, the hierarchy of Yukawa couplings is a problem in the SM without a compelling explanation. Another puzzle of the SM is that it does not give a reason of why there are only three generations of fermions.

Within this point of view, theories having a $SU(3)_C \times SU(3)_L \times U(1)_X $ gauge symmetry \cite{Georgi:1978bv,Singer:1980sw,Valle:1983dk,CarcamoHernandez:2017kra}
(called 3-3-1 models for short) can address many inexplicable issues of the
SM because those models have the following features: within
the QCD asymptotic freedom, the number of fermion generations is exactly three, the large mass splitting the heaviest quark and the two
 lighter ones is caused by quark family discrimination, the quantization of the electric charge \cite{deSousaPires:1998jc,VanDong:2005ux}
  and the CP violation \cite%
{Montero:1998yw,Montero:2005yb} are clarified in these models.
In addition, these theories contain a Peccei-Quinn symmetry, thus allowing to address the
strong-CP problem \cite{Pal:1994ba,Dias:2002gg,Dias:2003zt,Dias:2003iq}. Finally, the 3-3-1
models with heavy sterile neutrinos in the fermionic spectrum have cold dark matter candidates as
 weakly interacting massive particles (WIMPs) as shown in Refs. \cite%
{Mizukoshi:2010ky,Dias:2010vt,Alvares:2012qv,Cogollo:2014jia}.

In addition, discrete symmetry groups associated with the SM are an useful tool
 to explain the pattern of SM fermion masses and mixing angles. In particular the $\De (27)$ \cite
{Branco:1983tn,deMedeirosVarzielas:2006fc,Ma:2007wu,Bazzocchi:2009qg,Varzielas:2012nn,Bhattacharyya:2012pi,Ma:2013xqa,Nishi:2013jqa,Varzielas:2013sla,Aranda:2013gga,Ma:2014eka, Abbas:2014ewa,Abbas:2015zna,Varzielas:2015aua,Bjorkeroth:2015uou,Chen:2015jta,Vien:2016tmh,Hernandez:2016eod,CarcamoHernandez:2017owh,deMedeirosVarzielas:2017sdv,Bernal:2017xat,CarcamoHernandez:2018djj} discrete group has attracted a lot of attention since it provides a viable and very predictive description of the observed pattern of SM fermion masses and mixing angles.

In this work we build a $\De ( 27) $ flavor 3-3-1 model, where
the $\De ( 27) $ discrete symmetry is supplemented by the $%
Z_3 \times Z_{16}$ discrete group, providing a framework capable of reproducing the SM fermion masses and mixings. The model is much
 more economical in terms of scalar fields, discrete symmetries and number of scales than all 3-3-1 models with discrete symmetries proposed in the literature \cite{CarcamoHernandez:2017kra}.
In this model, the SM charged fermion masses and quark mixing angles are originated from the
spontaneous breakdown of the $\De (27) \times Z_3 \times Z_{16}$ discrete
symmetry and the masses for the light active neutrinos are produced by an inverse seesaw
mechanism, which can be probed at the LHC since the sterile neutrinos have masses at the TeV scale and can be produced via a Drell-Yan
 mechanism mediated by a $Z^{\prime}$ gauge boson. This neutrino mass generation mechanism is {\it not} presented in our previous 3-3-1 models
  with $\De(27)$ discrete symmetry \cite{Vien:2016tmh,Hernandez:2016eod} where the masses for the light active neutrinos are
  generated from a combination of type I and type II seesaw mechanisms \cite{Vien:2016tmh} and from a double seesaw
   mechanism  \cite{Hernandez:2016eod}. In those models the sterile neutrinos have extremely large masses, very much outside the
   LHC reach and the scalar spectrum,
  symmetries and number of scales are significantly much larger than in the current 3-3-1 model. Thus, this work corresponds to
  the first $\De(27)$ flavor 3-3-1 model with low scale seesaw mechanism.
The layout of the remainder of the paper is as follows. In section \ref{model} we
describe the model. Section \ref{quarksector} is devoted to
the implications of our model in quark masses and mixings. Section \ref%
{leptonsector} deals with lepton masses and mixings. We conclude in section %
\ref{conclusions}.

\section{The model}
\label{model}
The model considered in this work is based on the extended gauge symmetry $SU(3)_C\times SU(3)_L\times U(1)_X$, which is supplemented
by the $U(1)_{L_g}$ global lepton number symmetry and the $\De (27) \times Z_3\times Z_{16}$ discrete group. The $U(1)_{L_g}$ global
 lepton number symmetry, assumed to
be spontaneously broken by a vacuum expectation value (VEV) of a gauge-singlet scalar $\va$ to be introduced below. The $U(1)_{L_g}$
global lepton number symmetry is spontaneously broken down to a residual discrete $Z^{(L_g)}_2 $ lepton number symmetry under which
 the leptons are charged and the other particles are neutral. This residual discrete $Z^{(L_g)}_2$ lepton number symmetry prevents interactions
  with an odd number of leptons, thus forbidding proton decay. The corresponding massless Goldstone boson, Majoron, is phenomenologically
   harmless since it is a scalar singlet. In addition, we further assume that the $\De (27) \times Z_3\times Z_{16}$ discrete group is
   spontaneously broken as well. We introduce the $\De (27) $ and $Z_3$ discrete groups in order to reduce the number
    of model parameters, thus increasing the predictability of the model. The spontaneous breaking of the $\De (27) \times Z_3\times Z_{16}$ discrete
    symmetry produces the current pattern of SM fermion masses and mixing angles. In order to build the Yukawa terms invariant under
    all the symmetries of the model, we need to enlarge the scalar sector of the 3-3-1 model to include fourteen gauge singlet scalars.
    The electric charge of our model reads:
\[
Q = T_3-\fr{T_8}{\sqrt{3}}+X \, .
\]
We choose this kind of model (without non SM electric charges) in order to implement an inverse seesaw mechanism for the generation
 of the light neutrino masses and to avoid having in the fermion spectrum non SM fermions with exotic electric charges.
 Let us note that, in order to implement an inverse seesaw mechanism to generate the masses for the light active neutrinos,
 the fermion sector of the 3-3-1 model is expanded by adding three gauge singlets right handed Majorana neutrinos.

The full symmetry group $\mathcal{G}$ exhibits the following spontaneous breaking  pattern:
\bea
&&\mathcal{G}=SU(3)_C\times SU(3) _L\times U(1) _X\times U(1)_{L_g}\times \De (27) \times Z_3\times Z_{16}{%
\xrightarrow{\La _{int}}}\crn
&&\hspace{7mm}SU(3)_C\times SU(3) _L\times U(
1) _X\times U(1)_{L_g}{\xrightarrow{w,v_{\va}}}SU(3)_C\times SU(2)
_L \times U(1) _Y\times Z^{(L_g)}_2{\xrightarrow{u,v}}  \nn
\\
&&\hspace{7mm}SU(3)_C\times U(1) _Q\times Z^{(L_g)}_2,  \label{Group}
\eea
being $\La _{int}\gg w, v_{\va }\gg u,v.$

Moreover, the lepton number operator is defined as:
\be
L=\fr{4 \, T_8}{\sqrt{3}}+L_g\, ,
\label{Leptonnumber}
\ee
where the fact that the element at the bottom of the lepton triplet carries lepton number equal to $-1$, has been accounted for. Note that $L_g$ is a conserved charge associated with the $U(1)_{L_g}$ global symmetry and is interpreted as the ordinary lepton number.

Assignments of scalars under the $SU(3)_C\times SU(3)_L\times U(1)_X$ group and the
 $U(1)_{L_g}\times\De (27) \times Z_3\times Z_{16}$ fermionic assignments are displayed in Table \ref{tab:fermions} and \ref{tab:scalars}, respectively. For the quantum numbers of fermionic fields under the $SU(3)_C \times SU(3)_L \times U(1)_X$ symmetry, the reader is referred to Refs. \cite{Vien:2016tmh} and \cite{Hernandez:2016eod}.

The scalar triplets in this model are decomposed as:
\bea
\chi & = &%
\left( \chi _1^0  \, ,
\chi _2^-  \, ,
\fr 1{\sqrt{2}}(w+R _\chi \pm i I_\chi )\right)^T%
\, , \hs \phi =%
\left(\phi _1^{+} \, ,
\fr 1{\sqrt{2}}(v+R_{\phi}\pm i I_\phi ) \, ,
\phi_3^+\right)^T\, ,%
\crn
\eta & = &
\left(\fr 1 {\sqrt{2}}(u+R _\eta \pm i I_\eta ) \, ,
\eta _2^- \, ,
\eta _3^0\right)^T\, .%
\eea
Let us remark that the masses of non SM fermions and gauge bosons arise after the $SU(3) _L \times U(1) _X $ gauge
 symmetry is spontaneously broken at the scale $w$ by the scalar triplet $\chi$, whereas the SM particles get their masses after
  the spontaneous breaking of the SM electroweak gauge group, caused by the remaining scalar triplets, i.e., $\eta $ and $\phi$, which
   acquire Fermi scale VEVs equal to $u$ and $v$, respectively. In addition, we have fourteen EW scalar singlets in the scalar spectrum.
   They are crucial to build the Yukawa terms invariant under $U(1)_{L_g}\times \De (27) \times Z_3\times Z_{16}$ symmetry, which
    give rise to predictive textures for the fermion sector consistent with low energy SM fermion flavor data.

\begin{table}[tbp]
\begin{tabular}{|c|c|c|c|c|c|c|c|c|c|}
\hline
& $\chi$ & $\eta$ & $\phi$ & $\si$ & $\va$ & $\xi$ & $\zeta$ & $\Phi$ & $\Theta$ \\ \hline
$SU(3)_C$ & $\mathbf{1}$ & $\mathbf{1}$ & $\mathbf{1}$ & $\mathbf{1}$ & $\mathbf{1}$ & $\mathbf{1}$ & $\mathbf{1}$ & $\mathbf{1}$ & $\mathbf{1}$ \\ \hline
$SU(3)_L$ & $\mathbf{3}$ & $\mathbf{3}$ & $\mathbf{3}$ & $\mathbf{1}$ & $\mathbf{1}$ & $\mathbf{1}$ & $\mathbf{1}$ & $\mathbf{1}$ & $\mathbf{1}$ \\ \hline
$U(1)_X$ & $-\fr 1 3$ & $-\fr 1 3$ & $\fr 2 3$ & $0$ & $0$ & $0$ & $0$ & $0$ & $0$ \\ \hline
$L_g$ & $\fr 4 3 $ & $-\fr 2 3 $ & $-\fr 2 3$ & $0$ & $2$
& $0$ & $0$ & $0$ & $0$ \\ \hline
$\De(27)$ & $\mathbf{1}_{\mathbf{0,0}}$ & $\mathbf{1}_{\mathbf{1,0}}$ & $%
\mathbf{1}_{\mathbf{2,0}}$ & $\mathbf{1}_{\mathbf{0,0}}$ & $\mathbf{1}_{%
\mathbf{0,0}}$ & $\mathbf{3}$ & $\overline{\mathbf{3}}$ & $\overline{\mathbf{3}}$ & $\mathbf{3}$ \\ \hline
$Z_3$ & $0$ & $0$ & $0$ & $0$ & $0$ & $0$ & $1$ & $1$ & $-1$ \\ \hline
$Z_{16}$ & $0$ & $0$ & $0$ & $-1$ & $0$ & $0$ & $8$ & $8$ & $0$ \\ \hline
\end{tabular}%
\caption{Assignments of scalars under $SU(3)_C\times SU(3)_L\times U(1)_X\times U(1)_{L_g}\times \De (27) \times Z_3\times Z_{16}$.}
\label{tab:scalars}
\end{table}


\begin{table}[tbp]
\begin{tabular}{|c|c|c|c|c|c|c|c|c|c|c|c|c|c|c|c|c|c|}
\hline
		& $q_{1L}$ & $q_{2L}$ & $q_{3L}$ & $u_{1R}$ & $u_{2R}$ & $u_{3R}$ & $T_R$ & $%
d_{1R}$ & $d_{2R}$ & $d_{3R}$ & $D_{1R}$ & $D_{2R}$ & $l_L$ & $N_R$ & $%
e_{1R} $ & $e_{2R}$ & $e_{3R}$ \\ \hline
$L_g$ & $\fr 2 3 $ & $\fr 2 3 $ & $-\fr 2 3 $ & $0$ & $0$ &
$0$ & $-2$ & $0$ & $0$ & $0$ & $2$ & $2$ & $\fr 1 3$ & $-1$ & $1$ & $1$
& $1$ \\ \hline$$
$\De(27)$ & $\mathbf{1}_{\mathbf{0,0}}$ & $\mathbf{1}_{\mathbf{0,0}}$ & $%
\mathbf{1}_{\mathbf{0,0}}$ & $\mathbf{1}_{\mathbf{2,0}}$ & $\mathbf{1}_{%
\mathbf{2,0}}$ & $\mathbf{1}_{\mathbf{2,0}}$ & $\mathbf{1}_{\mathbf{0,0}}$ &
$\mathbf{1}_{\mathbf{1,0}}$ & $\mathbf{1}_{\mathbf{1,0}}$ & $\mathbf{1}_{%
\mathbf{1,0}}$ & $\mathbf{1}_{\mathbf{0,0}}$ & $\mathbf{1}_{\mathbf{0,0}}$ &
$\mathbf{3}$ & $\mathbf{3}$ & $\mathbf{1}_{\mathbf{0,0}}$ & $\mathbf{1}_{%
\mathbf{2,0}}$ & $\mathbf{1}_{\mathbf{1,0}}$ \\ \hline
$Z_3$ & $0$ & $0$ & $0$ & $0$ & $0$ & $0$ & $0$ & $0$ & $0$ & $0$ & $0$ & $%
0$ & $-1$ & $0$ & $0$ & $0$ & $0$ \\ \hline
$Z_{16}$ & $-2$ & $-1$ & $0$ & $4$ & $3$ & $0$ & $0$ & $5$ & $4$ & $3$ & $-2$
& $-1$ & $4$ & $4$ & $-4$ & $8$ & $6$ \\ \hline
\end{tabular}%
\caption{Assignments of fermions under $U(1)_{L_g}\times\De (27) \times Z_3\times Z_{16}$.}
\label{tab:fermions}
\end{table}

The quark and lepton Yukawa terms consistent with the symmetries of the model are given by:

\bea
-\mathcal{L}_Y^{(q) } &=&y_{11}^{(u) }\overline{q}%
_L ^{1}\phi ^{\ast }u_{1R}\fr{\si ^6 }{\La ^6 }%
+y_{12}^{(u)}\overline{q}_L ^{1}\phi ^{\ast }u_{2R}\fr{%
\si ^{5}}{\La ^{5}}+y_{21}^{(u)}\overline{q}_L ^2 \phi
^{\ast }u_{1R}\fr{\si ^{5}}{\La ^{5}}+y_{22}^{(u)}%
\overline{q}_L^2 \phi ^{\ast }u_{2R}\fr{\si ^4 }{\La ^4 }
\crn
&&+y_{13}^{(u)}\overline{q}_L ^{1}\phi ^{\ast }u_{3R}\fr{%
\si ^2 }{\La ^2 }+y_{31}^{(u)}\overline{q}_L ^3 \eta
u_{1R}\fr{\si ^4 }{\La ^4 }+y_{23}^{(u)}\overline{%
q}_L ^2 \phi ^{\ast }u_{3R}\fr{\si }{\La }+y_{32}^{\left(
u\right) }\overline{q}_L ^3 \eta u_{2R}\fr{\si ^3 }{\La ^3 }
\crn
&&+y_{33}^{(u)}\overline{q}_L ^3 \eta u_{3R}+y^{(
T) }\overline{q}_L ^3 \chi T_{R}+y_{1}^{(D) }\overline{q}%
_L ^{1}\chi ^{\ast }D_{1R}+y_2 ^{(D) }\overline{q}%
_L ^2 \chi ^{\ast }D_{2R}+y_{33}^{(d) }\overline{q}%
_L ^3 \phi d_{3R}\fr{\tau ^3 }{\La ^3 }\crn
&&+y_{11}^{(d) }\overline{q}_L ^{1}\eta ^{\ast }d_{1R}\fr{%
\si ^{7}}{\La ^{7}}+y_{12}^{(d) }\overline{q}_L ^{1}\eta
^{\ast }d_{2R}\fr{\si ^6 }{\La ^6 }+y_{21}^{(d) }%
\overline{q}_L ^2 \eta ^{\ast }d_{1R}\fr{\si ^6 }{\La ^6 }%
+y_{22}^{(d) }\overline{q}_L ^2 \eta ^{\ast }d_{2R}\fr{%
\si ^{5}}{\La ^{5}}\crn
&&+y_{13}^{(d) }\overline{q}_L^{1}\eta ^{\ast }d_{3R}\fr{%
\si ^{5}}{\La ^{5}}+y_{31}^{(d) }\overline{q}_L ^3 \phi
d_{1R}\fr{\si ^{5}}{\La ^{5}}+y_{23}^{(d) }\overline{%
q}_L ^2 \eta ^{\ast }d_{3R}\fr{\si ^4 }{\La ^4 }%
+y_{32}^{(d) }\overline{q}_L ^3 \phi d_{2R}\fr{\si ^4 %
}{\La ^4}+H.c,  \label{lyq}
\eea%
\bea
-\mathcal{L}_Y^{(l) } &=&y_{\phi e}^{(l) }\left(
\overline{l}_L \phi \Theta \right) _{\mathbf{\mathbf{1}}_{0,0}}e_{1R}\fr{%
\si ^8 }{\La ^9}+y_{\phi \mu }^{(l) }\left( \overline{l%
}_L \phi \Theta \right) _{\mathbf{1}_{1,0}}e_{2R}\fr{\si ^4 }{%
\La ^{5}}+y_{\phi \tau }^{(l) }\left( \overline{l}_L \phi
\Theta \right) _{\mathbf{1}_{2,0}}e_{3R}\fr{\si ^2}{\La ^3 }
\crn
&&+y_{\chi }^{(l) }\left( \overline{l}_L \chi N_{R}\right) _{%
\mathbf{\mathbf{1}}_{0,0}}+\fr 1 2 y_{1N}\left( N_{R}\overline{N_{R}^C }%
\right) _{\mathbf{3s}_{1}}\xi \fr{\va \si ^8 }{\La ^9 }%
+y_{2N}\left( N_{R}\overline{N_{R}^C }\right) _{\mathbf{3s}_2 }\xi \fr{%
\va \si ^8 }{\La ^9 }\crn
&&+y^{(1)}_{\phi }\varep _{abc}\left( \overline{l}_L ^{a}\left(
l_L ^C \right) ^{b}\right) _{\mathbf{3a}}\left( \phi ^{\ast }\right) ^{c}%
\fr{\zeta }{\La }+y^{(2) }_{\phi }\varep _{abc}\left( \overline{l}_L ^{a}\left(
l_L ^C \right) ^{b}\right) _{\mathbf{3a}}\left( \phi ^{\ast }\right) ^{c}%
\fr{\Phi }{\La }+H.c,  \label{Lyl}
\eea%
being $y_{ij}^{(u,d) }$ ($i,j=1,2,3$), $y_{\phi e}^{(l) }$, $y_{\phi \mu }{(l) }$, $y_{\phi \tau }^{(l) }$, $y_\chi ^{(l) }$, $y_{1N}$,
 $y_{2N}$, $y^{(1)}_{\phi }$ and $y^{(2) }_{\phi }$ $\mathcal{O}(1)$ dimensionless couplings.

In addition to these terms, the symmetries unavoidably allow terms obtained when replacing $\Theta$ with $\Phi^*\fr{\si^4}{\La^4}$ and $\zeta^*\fr{\si^4}{\La^4}$ as well as $\zeta$ and $\Phi$ with $\Theta^*\frac{\left(\si^*\right)^4}{\La^4}$ in $\tciLaplace _{Y}^{( l) }$. The resulting additional terms are:
\bea
&&\left(\overline{l}_L\phi\Phi^*\right) _{\mathbf{\mathbf{1}}_{0,0}}e_{1R}\fr{\si ^{12}}{\La ^{13}},\hspace{1cm}\left( \overline{l%
}_L \phi\Phi^*\right) _{\mathbf{1}_{1,0}}e_{2R}\fr{\si ^8 }{\La ^9 },\hspace{1cm}y_{\phi \tau }^{(l) }\left( \overline{l}_L \phi\Phi^*\right) _{\mathbf{1}_{2,0}}e_{3R}\fr{\si ^6}{\La ^{7}},\\
&&\left(\overline{l}_L \phi\zeta^*\right) _{\mathbf{\mathbf{1}}_{0,0}}e_{1R}\fr{\si ^{12}}{\La ^{13}},\hspace{0.8cm}\left( \overline{l%
}_L \phi\zeta^*\right) _{\mathbf{1}_{1,0}}e_{2R}\fr{\si ^8 }{\La ^9 },\hspace{0.8cm}y_{\phi \tau }^{(l) }\left( \overline{l}_L \phi\zeta^*\right) _{\mathbf{1}_{2,0}}e_{3R}\fr{\si ^6}{\La ^{7}},\hspace{0.8cm}\varep _{abc}\left( \overline{l}_L^a
\left(
l_L^C\right) ^b\right) _{\mathbf{3a}}\left( \phi ^{\ast }\right) ^{c}%
\fr{\Theta^*\left(\si^*\right)^4}{\La^5}.\nn
\eea
These terms will generate very subleading corrections to the charged lepton and Dirac neutrino mass matrices. Let us note that the hierarchy in the VEVs of the gauge singlet scalars (to be specified below) appearing in the aforementioned charged lepton and Dirac neutrino Yukawa interactions, will allow us to safely neglect these strongly suppressed corrections, and thus we will not consider them in our analysis.

As seen from Table \ref{tab:scalars} and Eq. (\ref{Lyl}), the $Z_3$ discrete %
symmetry guarantees that: only the $\De (27) $ scalar
triplets $\zeta $ and $\Phi$ appear in the Dirac neutrino Yukawa interactions, the $\xi $ is the only $\De (27)$ scalar
 triplet that participates in some of the neutrino Yukawa
interactions involving the right handed Majorana neutrinos $N_{iR}$ ($i=1,2,3
$) and $\Theta$ is the only $\De (27)$ scalar triplet appearing in the charged lepton Yukawa terms. Due to the different $\De (27) $
charge assignments for the quark fields given in Table \ref{tab:fermions}, there is no mixing between the SM and the non SM quarks.
We remark that $Z_{16}$ is the smallest discrete symmetry permitting to build the Yukawa term
$\left( \overline{l}_L \phi \Theta \right) _{\mathbf{\mathbf{1}}%
_{0,0}}e_{1R}\fr{\si ^8}{\La ^9 }$, required to provide a natural explanation for the small value
of the electron mass, which is $\la ^9%
\fr{v}{\sqrt{2}}$ times a $\mathcal{O}(1)$ coupling, where $\la =0.225$ is one of the Wolfenstein parameters. Therefore, the hierarchy among
charged fermion masses and quark mixing angles is produced by the spontaneous breakdown of
the $\De (27)\times Z_3\times Z_{16}$ discrete group.
Given that in this scenario the quark masses are related with the quark
mixing parameters, the vacuum expectation values of the scalars $\si $, $\va $, $%
\Theta_j$, $\xi _j$, $\zeta _j$, $\Phi_j$ ($j=1,2,3$) are taken as:
\be
v_\va \ll v_\zeta\sim\la^2\La \ll v_\Phi \sim v_\Theta \sim v_\xi \sim v_\si \sim
\la \La .
\label{VEVsingletshierarchy}
\ee

On the other hand, as indicated by Table \ref{ta:scalars}, three scalar triplets ($\chi, \eta,\phi$) and two
scalar singlets ($\si, \va$) are assigned into $\De(27)$ singlets, whereas the twelve other singlets ($\xi_j, \zeta_j, \Phi_j, \Theta_j$) $(j=1,2,3)$
are accommodated into 4 $\De(27)$ triplets.
  Out of the 14 scalar singlets, only $\va$ is assumed to acquire a VEV around the TeV scale, whereas the remaining 13 scalar singlets
  get VEVs at very high energy scale. The role of the fourteen scalar singlets is explained as follows. The
singlet scalar $\si$ is required to trigger the spontaneous breaking of the $Z_{16}$ discrete symmetry that generates
 the current pattern of SM charged fermion masses and mixing angles. The scalar singlet $\va$ is introduced to write the
 right handed Majorana neutrino Yukawa terms invariant under the $U(1)_{L_g}$ global lepton number symmetry. Let us note that $\va$ is the only
     scalar singlet charged under the $U(1)_{L_g}$ lepton number symmetry. Consequently, the VEV of the
      gauge singlet scalar $\va$ breaks the $U(1)_{L_g}$ global lepton number symmetry thus generating right handed Majorana neutrino
      mass terms that violate the lepton number by two units. These right handed Majorana neutrino mass terms are crucial for the
       implementation of the inverse seesaw mechanism crucial to produce the masses for the light active neutrinos.
       The lightness of the right handed Majorana neutrinos, which mediate the inverse seesaw mechanism, is explained by
        the thirteen dimensional Yukawa interactions involving a pair of right handed Majorana neutrinos and the %
      singlet scalar fields $\si$, $\xi$ and $\va$. After the spontaneous breaking of the $U(1)_{L_g}\times \De (27) \times Z_3\times Z_{16}$
       symmetry takes place, small right handed Majorana neutrino masses of the order of $\la^9v_\va$ are generated, being $\la=0.225$
       one of the Wolfenstein parameters. For $v_{\va}\sim 1$ TeV, the right handed Majorana neutrino masses are of the order of $1$ MeV.
       In addition we need three $\De(27)$ triplets $SU(3)_L$ scalar singlets, namely, $\Theta$, $\zeta$, $\Phi$ and $\xi$ that only appear
       in the charged lepton, Dirac neutrino and right handed Majorana neutrino Yukawa interactions, respectively.
       These $\De(27)$ scalar triplets that spontaneously break the $\De(27)$ discrete group, are required to have different VEV
       patterns in order to yield leptonic mixing parameters concordant with current data of neutrino oscillation experiments.
    Hence, the VEV patterns for the $\De (27) $ triplet SM singlet scalars $\Theta $, $\xi $, $\zeta $ and $\Phi$ are taken as:
\be
\left\langle \Theta \right\rangle =\fr{v_\Theta }{\sqrt{3}}\left(
1,e^{i\al },e^{i\bet }\right) ,\ms \left\langle \xi
\right\rangle =\fr{v_{\xi }}{\sqrt{3}}( 1,1,1) ,\hspace{1cm}%
\left\langle \zeta \right\rangle =\fr{v_\zeta }{\sqrt{2}}(1,0,1),\ms \left\langle \Phi \right\rangle
=v_\Phi\left(0,1,0\right)  ,
\end{equation}%
which are consistent with the scalar potential minimization conditions, as explained in detail in Refs. \cite{Hernandez:2016eod,CarcamoHernandez:2017owh,Ivanov:2014doa,deMedeirosVarzielas:2017glw}.

\section{Quark masses and mixings}
\label{quarksector}
The quark Yukawa interactions render the SM mass matrices for quarks:%
\be
M_{U}=\left(
\begin{array}{ccc}
a_{11}^{(u)}\la ^6  & a_{12}^{(u)}\la ^5
& a_{13}^{(u)}\la ^2  \\
a_{21}^{(u)}\la ^{5} & a_{22}^{(u)}\la ^4
& a_{23}^{(u)}\la \\
a_{31}^{(u)}\la ^4  & a_{32}^{(u)}\la ^3
& a_{33}^{(u)}%
\end{array}%
\right) \fr{v_{EW}}{\sqrt{2}},\ms \ms M_{D}=\left(
\begin{array}{ccc}
a_{11}^{(d) }\la ^7 & a_{12}^{(d) }\la ^6
& a_{13}^{(d) }\la ^5 \\
a_{21}^{(d) }\la ^6 & a_{22}^{(d) }\la ^5
& a_{23}^{(d) }\la ^4 \\
a_{31}^{(d) }\la ^{5} & a_{32}^{(d) }\la ^4
& a_{33}^{(u)}\la^3
\end{array}%
\right) \fr{v_{EW}}{\sqrt{2}},  \label{Mq}
\ee
where $\la =0.225$, $v_{EW}=246$ GeV and $a_{ij}^{(u,d) }$ ($%
i,j=1,2,3$) are dimensionless quantities of order unity, whose corresponding expressions are:
\bea
a_{nj}^{(U) } &=&y_{nj}^{(u)}\fr{v}{v_{EW}},%
\ms a_{3j}^{(u)}=y_{3j}^{(u)}\fr{u}{v_{EW}},\ms n=1,2,\crn
a_{nj}^{(d) } &=&y_{nj}^{(d) }\fr{u}{v_{EW}},%
\ms a_{3j}^{(d) }=y_{3j}^{(d) }\fr{v}{v_{EW}},\ms j=1,2,3.  \label{aq}
\eea
Moreover, the different $\De (27) $ charge assignments for the quark fields produces the absence of mixings between exotic quarks and SM quarks.
The masses of the exotic quarks are:
\be
m_{T}=y^{(T) }\fr{w}{\sqrt{2}},\ms
m_{D_{n}}=y_{n}^{(D) }\fr{w}{\sqrt{2}},\ms
n=1,2. \label{mexotics}
\ee
Considering that the spontaneous breakdown of the $\De (27) \times Z_3 \times Z_{16}$ discrete group produces the observed
 pattern of charged fermion mass and quark mixing angles and for the sake of simplicity, we take a benchmark scenario characterized by the relations:
\bea
a_{12}^{(u)} &=&a_{21}^{(u)},\hspace{0.7cm}%
a_{31}^{(u)}=y_{13}^{(u)},\hspace{0.7cm}%
a_{32}^{(u)}=y_{23}^{(u)},\crn
a_{12}^{(d) } &=&\left\vert a_{12}^{(d) }\right\vert
e^{-i\tau _{1}},\hspace{0.75cm}a_{21}^{(d) }=\left\vert
a_{12}^{(d) }\right\vert e^{i\tau _{1}}, \\
a_{13}^{(d) } &=&\left\vert a_{13}^{(d) }\right\vert
e^{-i\tau _2 },\hspace{0.75cm}a_{31}^{(d) }=\left\vert
a_{13}^{(d) }\right\vert e^{i\tau _2},\hspace{0.75cm}%
a_{23}^{(d) }=a_{32}^{(d) }.  \nn
\eea%
Furthermore, motivated by naturalness arguments, we set $a_{33}^{(u)}=1$. Then, the experimental values of the quark masses,
 mixing angles and CP violating phase can be well reproduced for the following benchmark point:
\bea
a_{11}^{(u)} &\simeq &0.58,\ms a_{22}^{(u)
}\simeq 2.19,\ms a_{12}^{(u)}\simeq 0.67,\crn
a_{13}^{(U) } &\simeq &0.80,\ms a_{23}^{(U)
}\simeq 0.83,\ms a_{11}^{(d) }\simeq 1.96,\crn
a_{12}^{(d) } &\simeq &0.53,\ms a_{13}^{(d)
}\simeq 1.07,\ms a_{22}^{(d) }\simeq 1.93, \\
a_{23}^{(d) } &\simeq &1.36,\hspace{0.5cm}a_{33}^{\left(
D\right) }\simeq 1.35,\hspace{0.5cm}\tau _{1}\simeq 9.56^{\circ },\hspace{%
0.5cm}\tau _2 \simeq 4.64^{\circ }.  \notag
\eea%

\begin{table}[tbh]
\bc
\begin{tabular}{c|l|l|}
\hline\hline
Observable & Model value & Experimental value \\ \hline
$m_u(MeV)$ & \quad $1.44$ & \quad $1.45_{-0.45}^{+0.56}$ \\ \hline
$m_c(MeV)$ & \quad $656$ & \quad $635\pm 86$ \\ \hline
$m_t(GeV)$ & \quad $177.1$ & \quad $172.1\pm 0.6\pm 0.9$ \\ \hline
$m_d(MeV)$ & \quad $2.9$ & \quad $2.9_{-0.4}^{+0.5}$ \\ \hline
$m_s(MeV)$ & \quad $57.7$ & \quad $57.7_{-15.7}^{+16.8}$ \\ \hline
$m_b(GeV)$ & \quad $2.82$ & \quad $2.82_{-0.04}^{+0.09}$ \\ \hline
$\sin \theta _{12}$ & \quad $0.225$ & \quad $0.225$ \\ \hline
$\sin \theta _{23}$ & \quad $0.0412$ & \quad $0.0412$ \\ \hline
$\sin \theta _{13}$ & \quad $0.00351$ & \quad $0.00351$ \\ \hline
$\de $ & \quad $64^{\circ }$ & \quad $68^{\circ }$ \\ \hline\hline
\end{tabular}%
\ec
\caption{Model and experimental values related to the quark masses and CKM
parameters.}
\label{Tab}
\end{table}
As displayed in Table \ref{Tab}, the resulting physical quark mass spectrum \cite{Bora:2012tx,Xing:2007fb}, mixing angles and
CP violating phase \cite{Olive:2016xmw} obtained in our model, are concordant with the low energy quark flavor data.

Hereafter we briefly discuss an effect of quarks on flavor changing processes in our model. The absence of mixings between the
SM and exotic quarks, which arises from the $\De(27)$ symmetry, leads to the fact that the exotic fermions will not exhibit flavor
 changing decays into SM quarks and gauge (or Higgs) bosons.
  After being pair produced the exotic fermions will decay into the SM quarks and the intermediate states of heavy gauge
bosons, which in turn decay into the pairs of the SM fermions, see e.g. \cite
{Cabarcas:2008ys}. The present lower bounds on the $Z^{\prime }$ gauge boson mass in 3-3-1 models resulting from LHC searches,
reach around $2.5$ \text{TeV} \cite{Salazar:2015gxa}. These limits generate a
 bound of about $6.3$ TeV for the 3-3-1 gauge symmetry breaking scale $w$. In addition, lower limits on the $Z^\prime $ gauge
  boson mass varying from $1$ TeV up to $3$ TeV can be obtained by using the electroweak data associated with the decays
  $B_{s,d}\rightarrow \mu ^+\mu ^-$ and $B_{d}\rightarrow K^{\ast }(K)\mu ^+\mu ^-$ \cite{CarcamoHernandez:2005ka,Martinez:2008jj,Buras:2013dea,Buras:2014yna,Buras:2012dp}. The exotic quarks can be pair
   produced at the LHC via Drell-Yan and gluon
fusion processes mediated by charged gauge bosons and gluons, respectively. A detailed study of the exotic quark production
 at the LHC and the exotic quark decay modes is beyond the scope of this work and will be done elsewhere.

\section{Lepton masses and mixings}
\label{leptonsector}
Using the charged lepton Yukawa interactions we obtain the mass matrix for charged leptons:
\be
M_l=R_{lL}\textrm{diag}\left( m_e,m_\mu ,m_\tau \right) ,\ms R_{lL}=%
\fr{1}{\sqrt{3}}\left(
\begin{array}{ccc}
1 & 0 & 0 \\
0 & e^{i\al } & 0 \\
0 & 0 & e^{i\bet}%
\end{array}%
\right) \left(
\begin{array}{ccc}
1 & 1 & 1 \\
1 & \om ^2 & \om  \\
1 & \om  & \om ^2%
\end{array}%
\right) , \ms \om =e^{\fr{2\pi i} 3},  \label{Uclep}
\end{equation}%
with the masses of the charged leptons determined as
\be
m_e=a_{1}^{(l) }\la ^9\fr{v_{EW}}{\sqrt{2}},\hspace{1cm}%
m_\mu =a_2^{(l) }\la ^5\fr{v_{EW}}{\sqrt{2}},\hspace{1cm}%
m_\tau =a_3^{(l) }\la ^3\fr{v_{EW}}{\sqrt{2}},
\label{leptonmasses}
\ee
being $a_i^{(l) }$ ($i=1,2,3$) $\mathcal{O}(1)$ dimensionless quantities.

In addition, with the help of Eq. (\ref{Lyl}), the following expressions for the neutrino mass terms are obtained:
\be
-\mathcal{L}_{mass}^{(\nu) }=\fr 1 2 \left(
\begin{array}{ccc}
\overline{\nu _L^C} & \overline{\nu _R} & \overline{N_R}%
\end{array}%
\right) M_\nu \left(
\begin{array}{c}
\nu _L \\
\nu _R^C \\
N_R^C
\end{array}%
\right) +H.c.
\ee
with full $9\times 9$ mass matrix for the neutrino fields is given as:
\be
M_\nu =\left(
\begin{array}{ccccccccc}
 0 & 0 & 0 & 0 & \fr{v \om ^2 v_{\zeta } y_\phi }{2 \La } & -\frac{r v \om  v_\zeta  y_\phi }{2 \La } & 0 & 0 & 0 \\
 0 & 0 & 0 & -\fr{v \om ^2 v_\zeta  y_\phi }{2 \La } & 0 & \frac{v v_\zeta  y_\phi }{2 \La } & 0 & 0 & 0 \\
 0 & 0 & 0 & \fr{r v \om  v_{\zeta } y_{\phi }}{2 \La } & -\frac{v v_{\zeta } y_{\phi }}{2 \La } & 0 & 0 & 0 & 0 \\
 0 & -\fr{v \om ^2 v_{\zeta } y_{\phi }}{2 \La } & \fr{r v \om  v_\zeta  y_\phi }{2 \La } & 0 & 0 & 0 & \fr{w y_{\chi
   }}{\sqrt{2}} & 0 & 0 \\
 \frac{v \om ^2 v_\zeta  y_\phi }{2 \La } & 0 & -\fr{v v_\zeta  y_\phi }{2 \La } & 0 & 0 & 0 & 0 & \fr{w y_{\chi
   }}{\sqrt{2}} & 0 \\
 -\fr{r v \om  v_\zeta  y_\phi }{2 \La } & \fr{v v_\zeta  y_\phi }{2 \La } & 0 & 0 & 0 & 0 & 0 & 0 & \fr{w y_{\chi
   }}{\sqrt{2}} \\
 0 & 0 & 0 & \fr{w y_chi }{\sqrt{2}} & 0 & 0 & \fr{y_{1N}v_\xi  v_\si^8 v_\va}{\La^9} & \fr{y_{1N}x v_\xi  v_\si^8
   v_\va }{\La^9} & \fr{y_{1N}x v_\xi  v_\si^8 v_\va}{\La^9} \\
 0 & 0 & 0 & 0 & \fr{w y_\chi }{\sqrt{2}} & 0 & \fr{y_{1N}x v_\xi  v_\si^8 v_\va }{\La^9} & \fr{y_{1N}v_\xi  v_\si^8
   v_\va }{\La^9} & \fr{y_{1N}x v_\xi  v_\si^8 v_\va }{\La^9} \\
 0 & 0 & 0 & 0 & 0 & \fr{w y_\chi }{\sqrt{2}} & \fr{y_{1N}x v_\xi  v_\si^8 v_\va }{\L^9} & \fr{y_{1N}x v_\xi  v_\si^8 v_\va}{\La^9} & \fr{y_{1N}v_\xi  v_\si^8 v_\va }{\La^9} \\
\end{array}
\right),
\ee
where
\bea
\om =e^{\fr{2\pi i}{3}},\ms
x=\fr{y_{2N}%
}{y_{1N}}\, ,  \ms r=\fr{v_\Phi}{v_\zeta}.
\eea
 Remember that the spectrum of the physical neutrino fields is formed by 3 light active neutrinos and 6 sterile exotic pseudo-Dirac
  neutrinos having masses of the order of $\sim
\pm w$ and a small mass difference of about $\fr{v_{\xi } v_\si^8 v_{\va } }{\La^9}$. The sterile neutrinos can be pair produced at the Large Hadron
Collider (LHC), via a Drell-Yan annihilation mediated by a heavy $Z^\prime $ gauge boson. The mixings of these sterile neutrinos
with the SM neutrinos allow the former to decay into
SM particles, so that the final decay products will be a
 SM charged lepton and a $W$ gauge boson. Hence, observing an excess of events in the dilepton
final states above the
SM background at the LHC, might be a signature concordant with this model.
Studies of inverse seesaw neutrino signatures at the Large Hadron Collider and International Linear Collider as well as the
 production of heavy neutrinos at the LHC are carried out in Refs. \cite{Das:2012ze,Das:2016hof}. A comprehensive study
 of the implications of our model at colliders goes out of the purpose of this work and will be done elsewhere.

After the implementation of the inverse seesaw mechanism, one finds the mass matrix for the light active neutrino fields:
\be
M_\nu ^{(1) }=z\left(
\begin{array}{ccc}
-1 +i\sqrt{3}-(1+i\sqrt{3}) r^2 - 4 r x& (1-i\sqrt{3}) r - 2(r-i\sqrt{3}) x &(1+i\sqrt{3})[1 + r(1 + r) x] \\
 (1-i\sqrt{3}) r - 2(r-i\sqrt{3}) x & (1+i\sqrt{3})(1+2x) &-2 r - (3 + r) x + i \sqrt{3}(r-1)x \\
(1+i\sqrt{3})\left[1 + r(1 + r) x\right] &-2 r - (3 + r) x + i \sqrt{3}(r-1)x  &2+2(r-i\sqrt{3}r)x-(1+i\sqrt{3}) r^2
\end{array}\right) ,
\ee
with
\be
z=\fr{y_\phi^2 y_{1N} v^2  v^2_\zeta v_{\xi} v_\va v_\si^8}{4\La^{11} w^2 \left(y^{(l)}_\chi\right)^2}.
\ee
Thus, small masses for active neutrinos are naturally produced in our model because these masses are inversely proportional to powers of the large
 model cutoff  $\La $ and feature a quadratic dependence on the very small VEVs of the $SU(3)_L $ singlet and $\De(27)$ triplet
  scalar fields $\zeta$ and $\Phi$. On the other hand, from the VEV hierarchy of Eq. (\ref{VEVsingletshierarchy}) and assuming
  $v_{\va}\sim 1$ TeV, we notice that $z\sim \la^{20}v_{\va}\sim 10^{-13}$ TeV$=$$0.1$ eV, which is associated
   with the light active neutrino mass scale. Thence, the small value of the active neutrino mass scale is naturally explained in our model.

With the help of the rotation matrix $R_{\nu }$, the mass matrix $M_{\nu }^{(1) }$ for the light active neutrinos is easily diagonalized:
\be
R_\nu ^{\dag}M_\nu ^{(1) }R_\nu =\left\{
\begin{array}{l}
\left(
\begin{array}{ccc}
0 & 0 & 0 \\
0 & m_2 & 0 \\
0 & 0 & m_3%
\end{array}%
\right) ,\hspace{0.5cm} R_\nu =\left( \begin{array}{ccc}
\fr{A_1}{\sqrt{1+|A_1|^2+|A_2|^2}}& \fr{B_1}{\sqrt{1+|B_1|^2+|B_2|^2}} &
\fr{C_1}{\sqrt{1+|C_1|^2+|C_2|^2}}\\ \fr{A_2}{\sqrt{1+|A_1|^2+|A_2|^2}}&
\fr{B_2}{\sqrt{1+|B_1|^2+|B_2|^2}} &
\fr{C_2}{\sqrt{1+|C_1|^2+|C_2|^2}}\\ \fr{1}{\sqrt{1+|A_1|^2+|A_2|^2}}&
\fr{1}{\sqrt{1+|B_1|^2+|B_2|^2}} & \fr{1}{\sqrt{1+|C_1|^2+|C_2|^2}}
\end{array}\right) ,\hspace{0.5cm}\mbox{for \ NH}\ \  \\
\left(
\begin{array}{ccc}
m_2 & 0 & 0 \\
0 & m_3 & 0 \\
0 & 0 & 0%
\end{array}%
\right) ,\hspace{0.5cm}R_{\nu }=\left( \begin{array}{ccc}
\fr{B_1}{\sqrt{1+|B_1|^2+|B_2|^2}} &
\fr{C_1}{\sqrt{1+|C_1|^2+|C_2|^2}}&\fr{A_1}{\sqrt{1+|A_1|^2+|A_2|^2}}\\
\fr{B_2}{\sqrt{1+|B_1|^2+|B_2|^2}} &
\fr{C_2}{\sqrt{1+|C_1|^2+|C_2|^2}}&\fr{A_2}{\sqrt{1+|A_1|^2+|A_2|^2}}\\
\fr{1}{\sqrt{1+|B_1|^2+|B_2|^2}}&
\fr{1}{\sqrt{1+|C_1|^2+|C_2|^2}}&\fr{1}{\sqrt{1+|A_1|^2+|A_2|^2}}
\end{array}\right) ,\hspace{0.5cm}\mbox{for \ IH}%
\end{array}%
\right.  \label{Rnu}
\ee
Here the following notations are used
\bea
A_1 &= &\fr 1 2 (-1+i\sqrt{3}), \hspace{0.5cm} A_2=-\fr 1 2 (1+i\sqrt{3}) ,\crn
B_1 &=&\fr{2(1+r+r^2)\{-(3+i\sqrt{3})x+r\left[-1-i\sqrt{3}+(1-i\sqrt{3})x\right]\}}{\{1-i\sqrt{3}+r[4i+(1+i\sqrt{3})]\}[r(1+i\sqrt{3}+2ix)+2\sqrt{3}x]}, \hspace{0.2cm} B_2=\fr{-3i+\sqrt{3}}{2i+(i+\sqrt{3})r},\crn
C_1 &= &1-\fr{2 \sqrt{3}}{\sqrt{3}+i(3+2r)}, \hspace{0.5cm} C_2=\fr 1{\fr 3 2 -i\fr{\sqrt{3}}{2}+r}.  \label{ABCi}
\eea
Thus, this model predicts in the physical spectrum of active neutrinos one massless neutrino and two active ones. Here NH and IH correspond to normal and inverted neutrino mass hierarchies, respectively and the light active neutrino masses $m_2$ and $m_3$ are given by:
\bea
 m_2&=&(1+i\sqrt{3})(1-r)[1+r+(3+r)x]z,\hs m_3=(1+i\sqrt{3})(1-r^2)(1-x)z,  \label{m2m3}
\eea
Now, taking into account the Eqs. (\ref{Uclep}) and (\ref{Rnu}), the Pontecorvo–Maki–Nakagawa–Sakata (PMNS) leptonic mixing matrix is given by:
\bea
&&U^N \equiv R_{lL}^\dag R_\nu \crn
 &&=\left(
\begin{array}{ccc}
\fr{(e^{-i\bet}+A_1+A_2 e^{-i\al})}{\sqrt{3}\Ga_{1}}&\fr{(e^{-i\beta}+B_1+B_2 e^{-i\al})}{\sqrt{3}\Ga_2 } & \fr{(e^{-i\beta}+C_1+C_2 e^{-i\al})}{\sqrt{3}\Ga_3 }\\
\fr{-(3i+\sqrt{3})e^{-i\beta}+2\sqrt{3}A_1+(3i-\sqrt{3})A_2e^{-i\al}}{6\Ga_{1}}& \fr{-(3i+\sqrt{3})e^{-i\beta}+2\sqrt{3}B_1+(3i-\sqrt{3})B_2e^{-i\al}}{6\Ga_2 } &  \fr{-(3i+\sqrt{3})e^{-i\beta}+2\sqrt{3}C_1+(3i-\sqrt{3})C_2e^{-i\al}}{6\Ga_3 }\\
\fr{(3i-\sqrt{3})e^{-i\beta}+2\sqrt{3}A_1-(3i+\sqrt{3})A_2e^{-i\al}}{6\Ga_{1}}&\fr{(3i-\sqrt{3})e^{-i\beta}+2\sqrt{3}B_1-(3i+\sqrt{3})B_2e^{-i\al}}{6\Ga_2 } &  \fr{(3i-\sqrt{3})e^{-i\beta}+2\sqrt{3}C_1-(3i+\sqrt{3})C_2e^{-i\al}}{6\Ga_3 }
\end{array}\right) \notag\\ \label{RnuN}
\eea
for Normal ordering, and
\bea
 &&U^I\equiv R_{lL}^\dag R_\nu \crn
 &&=
\begin{array}{l}
 \left(
\begin{array}{ccc}
\fr{(e^{-i\bet}+B_1+B_2 e^{-i\al})}{\sqrt{3}\Ga_2 } & \fr{(e^{-i\bet}+C_1+C_2 e^{-i\al})}{\sqrt{3}\Ga_3 }&\fr{(e^{-i\bet}+A_1+A_2 e^{-i\al})}{\sqrt{3}\Ga_{1}}\\
 \fr{-(3i+\sqrt{3})e^{-i\bet}+2\sqrt{3}B_1+(3i-\sqrt{3})B_2e^{-i\al}}{6\Ga_2 } &  \fr{-(3i+\sqrt{3})e^{-i\bet}+2\sqrt{3}C_1+(3i-\sqrt{3})C_2e^{-i\al}}{6\Ga_3 }&\fr{-(3i+\sqrt{3})e^{-i\bet}+2\sqrt{3}A_1+(3i-\sqrt{3})A_2e^{-i\al}}{6\Ga_{1}}\\
\fr{(3i-\sqrt{3})e^{-i\bet}+2\sqrt{3}B_1-(3i+\sqrt{3})B_2e^{-i\al}}{6\Ga_2 } &  \fr{(3i-\sqrt{3})e^{-i\bet}+2\sqrt{3}C_1-(3i+\sqrt{3})C_2e^{-i\al}}{6\Ga_3 }&\fr{(3i-\sqrt{3})e^{-i\bet}+2\sqrt{3}A_1-(3i+\sqrt{3})A_2e^{-i\al}}{6\Ga_{1}}
\end{array}\right)
\end{array}\nn
\label{RnuI}
\eea for Inverted ordering. Here the functions $\Ga_{1}$, $\Ga_2 $ and $\Ga_3 $ are defined as:
\bea
\Ga_1&=& \left(\fr{A_2B_1-A_1B_2+B_2C_1-B_1C_2+A_1C_2-A_2C_1}{B_2C_1-B_1C_2}\right)^{1/2} ,\crn
\Ga_2 &=&\left(\fr{A_2B_1-A_1B_2+B_2C_1-B_1C_2+A_1C_2-A_2C_1}{A_1C_2-A_2C_1}\right)^{1/2},\crn
\Ga_3&=& \left(\fr{A_2B_1-A_1B_2+B_2C_1-B_1C_2+A_1C_2-A_2C_1}{A_2B_1-A_1B_2}\right)^{1/2},  \label{Ga123}
\eea
where $A_i, B_i, C_i \hspace{0.1 cm} (i=1,2)$ are given in Eq. (\ref{ABCi}).

We point out that there are 8 effective free parameters ($a^{(l)}_{1,2,3}, r, x, z, \al, \beta$) to describe the lepton sector of this model.
These parameters can be adjusted to reproduce the experimental values of the eight physical observables in the lepton
sector, including 3 masses for the charged leptons, 2 neutrino mass squared differences and 3 leptonic mixing parameters. We obtain that the scenario of
inverted neutrino mass ordering of our model cannot be fitted to the neutrino oscillation experimental data, however,
the lepton sector parameters of the model under consideration are highly consistent with the recent experimental data in the
case of normal ordering. Indeed, in the Normal Hierarchy, with $A_i, B_i, C_i  (i=1,2) $ given by Eq. (\ref{ABCi})
and $R_{lL}$ in Eq. (\ref{Uclep}), the matrix $U^N$ in Eq. (\ref{RnuN}) depends on four parameters $\al,\bet, r$
and $x$, in which three elements $U^N_{11,21,31}$ in Eq. (\ref{RnuN}) depend only on two parameters $\al, \bet$,
three elements $U^N_{13,23,33}$ depend on three parameters $\al,\bet, r$ and three elements $U^N_{12,22,32}$
 depend on four parameters $\al,\bet, r $ and $x$.

\begin{figure}[h]
\bc
\includegraphics[width=11.5cm, height=10.0cm]{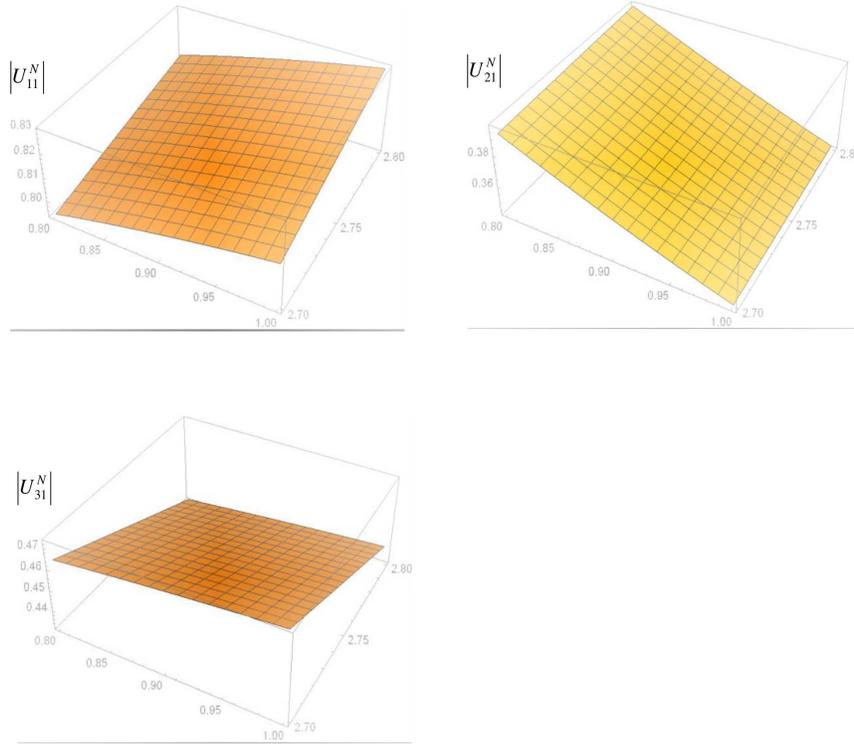}
\vspace*{-0.1cm} \caption[$U^N_{11,21,31}$ as functions of
 $\al, \bet$ with $\al \in (0.8, 1.0) rad$ and $\bet \in (2.7, 2.8) rad$ in the Normal Hierarchy]{$U^N_{11,21,31}$ as functions of
 $\al, \bet$ with $\al \in (0.8, 1.0) rad$ and $\bet \in (2.7, 2.8) rad$ in the Normal Hierarchy.}\label{Ui1N3d}
\vspace*{-0.1cm}

\ec
\end{figure}

\begin{figure}[h]
\bc
\includegraphics[width=12.0cm, height=11.0cm]{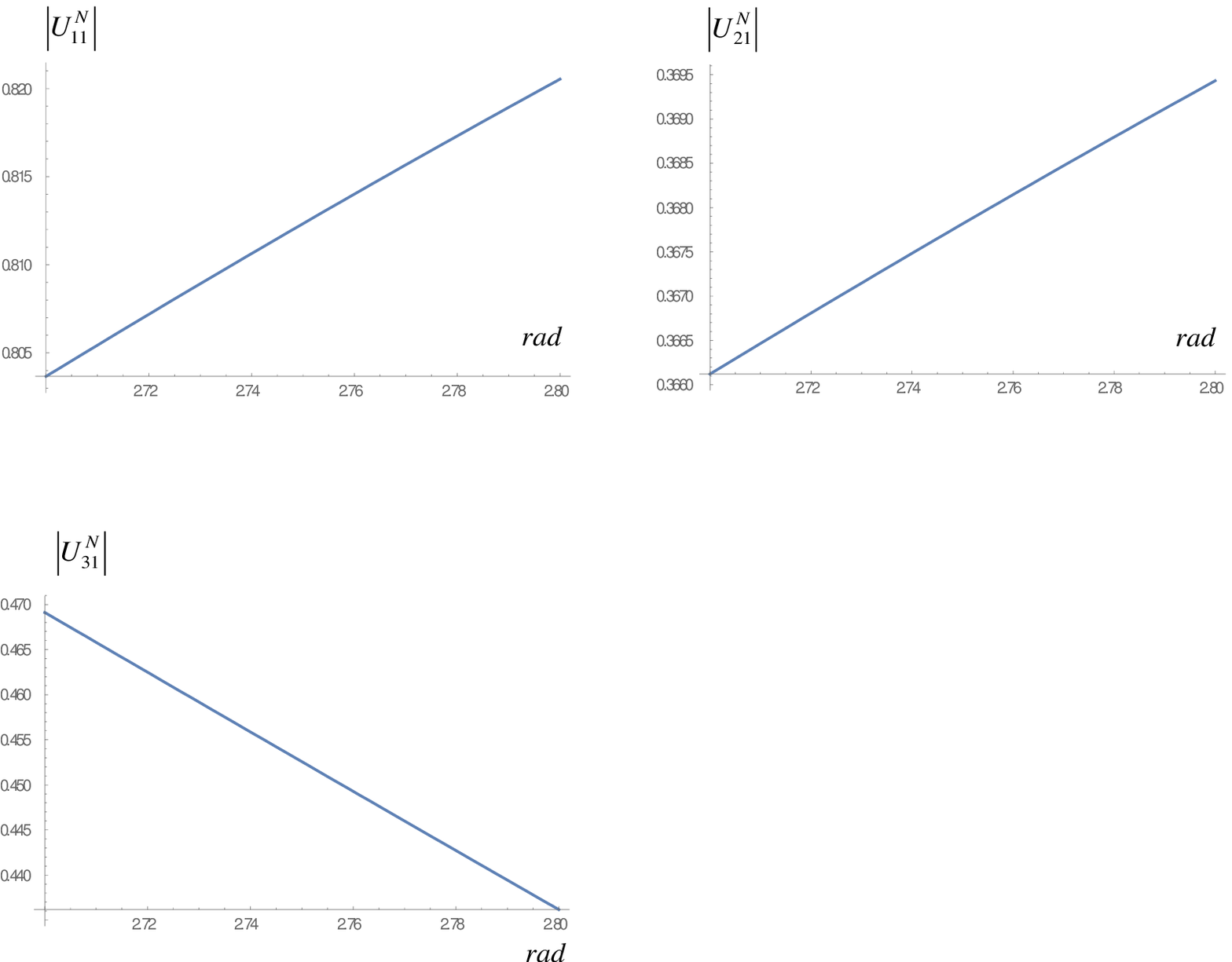}
\vspace*{-0.4cm} \caption[$U^N_{11,21,31}$ as functions of
 $\bet$ with $\bet \in (2.7, 2.8) rad$ and $\al=0.9 rad$ in the Normal Hierarchy]{$U^N_{11,21,31}$ as functions of
 $\bet$ with $\bet \in (2.7, 2.8) rad$ and $\al=0.9 rad$ in the Normal Hierarchy.}\label{Ui1N1d}
\vspace*{-0.3cm}

\ec
\end{figure}

\begin{figure}[h]
\bc
\includegraphics[width=12.0cm, height=11.0cm]{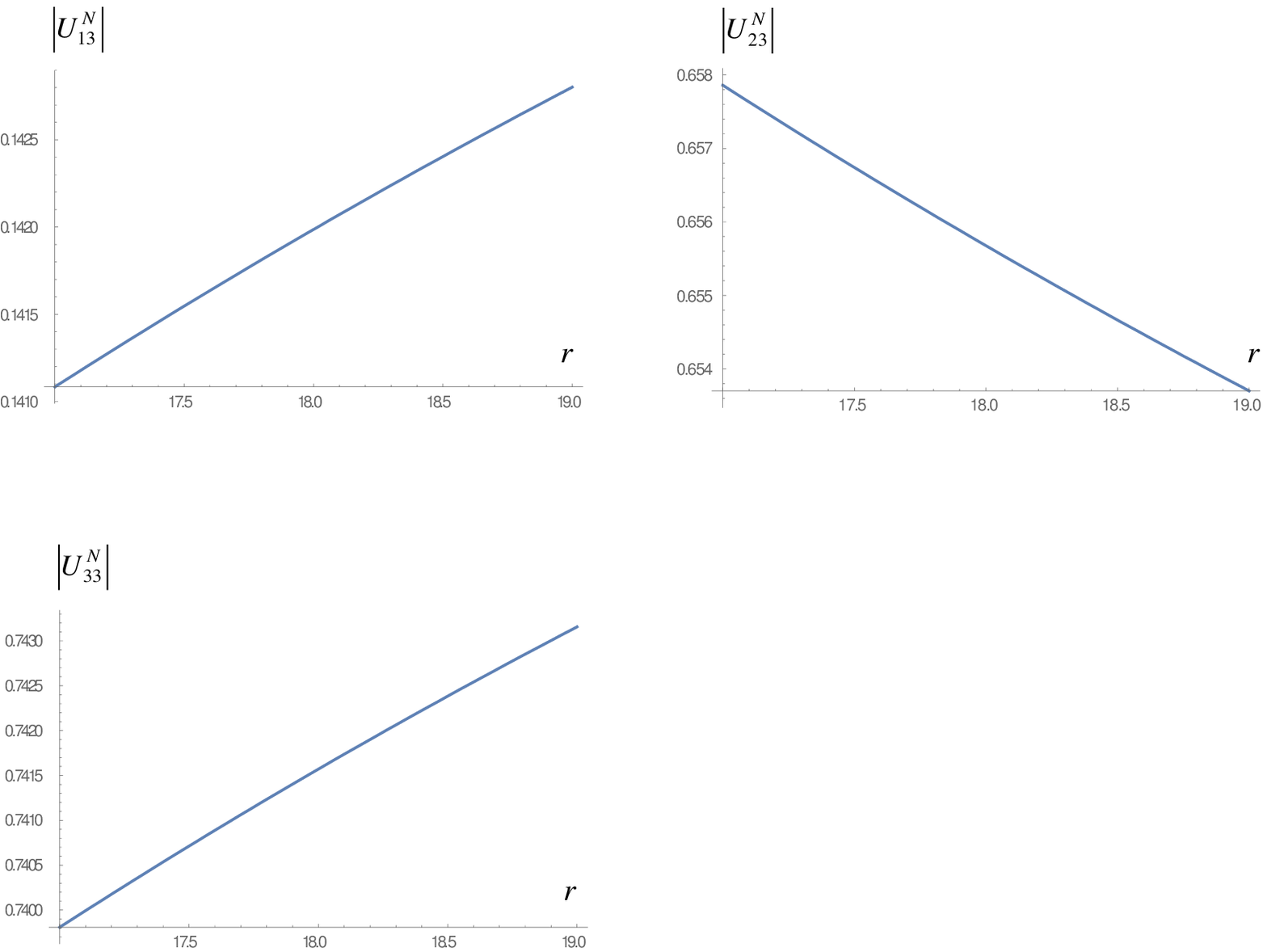}
\vspace*{-0.4cm} \caption[$U^N_{13,23,33}$ as functions of
 $r$ with $r \in (13.5, 14.0) $ in the Normal Hierarchy]{$U^N_{13,23,33}$ as functions of
 $r$ with $r \in (13.5, 14.0) $ in the Normal Hierarchy.}\label{Ui3N1d}
\vspace*{-0.3cm}

\ec
\end{figure}

\begin{figure}[h]
\bc
\includegraphics[width=12.0cm, height=11.0cm]{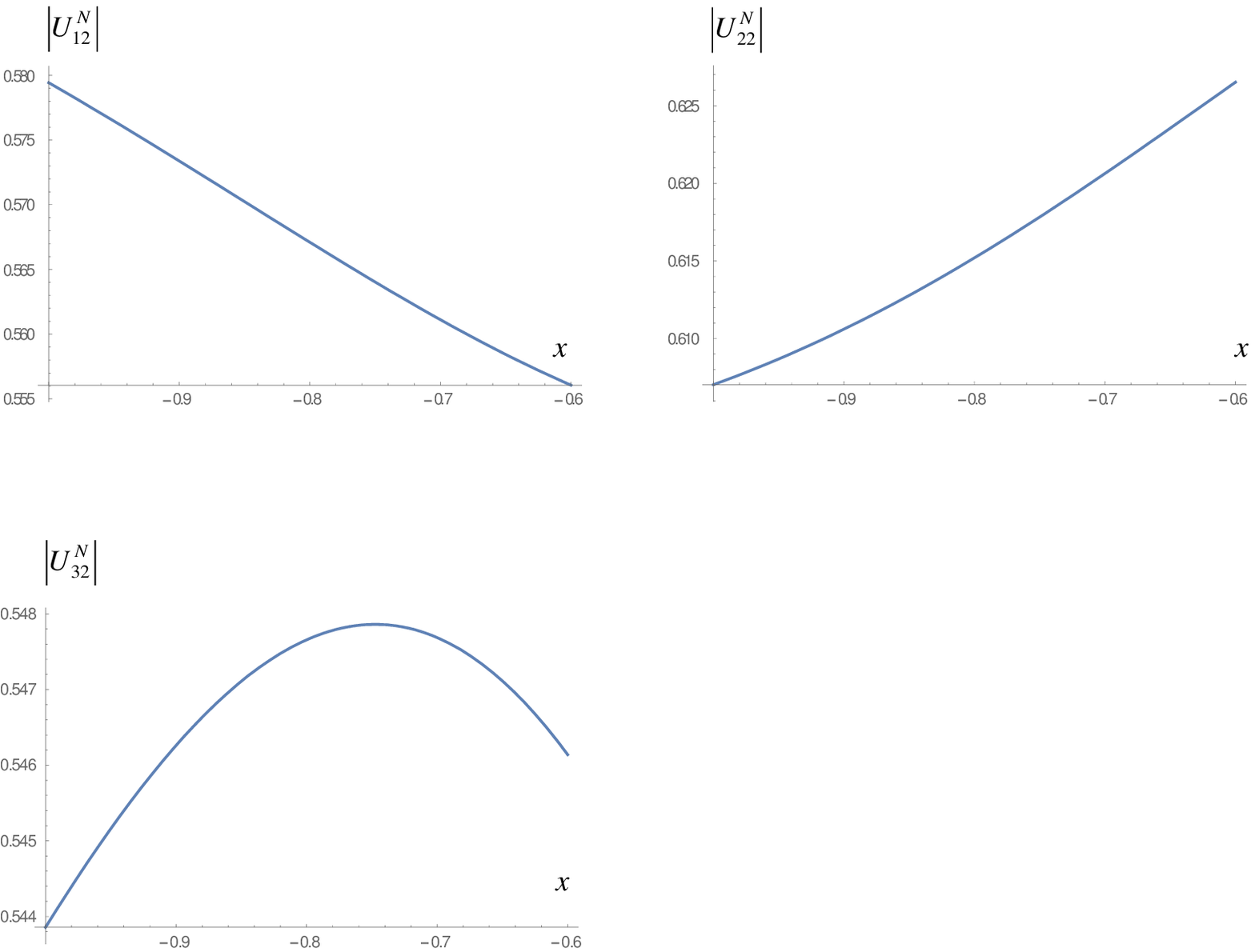}
\vspace*{-0.4cm} \caption[$U^N_{12,22,32}$ as functions of
 $x$ with $x \in (-1.0, -0.6) $ in the Normal Hierarchy]{$U^N_{12,22,32}$ as functions of
 $x$ with $x \in (-1.0, -0.6) $ in the Normal Hierarchy.}\label{Ui2N1d}
\vspace*{-0.3cm}

\ec
\end{figure}

In Fig. \ref{Ui1N3d}, we have plotted the
magnitudes of $U^N_{11,21,31}$ as functions of
 $\al, \bet$ with $\al \in (0.8, 1.0) \,\mathrm{rad}$ and $\bet \in (2.7, 2.8) \,\mathrm{rad}$.
If $\al=0.9 \,\mathrm{rad}$, the dependence of $U^N_{11,21,31}$ on $\beta$ with $\bet \in (2.7, 2.8) \,\mathrm{rad}$ is depicted in Fig. \ref{Ui1N1d}.
For the case $\bet=2.75 \,\mathrm{rad\,} (157.563^\circ)$ we get $U^N_{11}=0.812333, U^N_{21}=0.367816, U^N_{31}=0.452577$, as well as the following relations:
\bea
U^N_{13}&=& \fr{(0.233612 + 0.179369 i)+(0.043704 - 0.220352 i) r}{(1.5 - 0.866025 i+r)\sqrt{2+\fr{1}{3 + r (3 + r)}}} ,\crn
U^N_{23} &=& \fr{(1.68788 + 1.82961 i) + (0.653343 + 0.572327 i) r}{(1.5 - 0.866025 i+r)\sqrt{2+\fr{1}{3 + r (3 + r)}}},\crn
U^N_{33}&=& \fr{(0.67658 - 0.50898 i) + (1.035 - 0.351975 i) r}{(1.5 - 0.866025 i+r)\sqrt{2+\fr{1}{3 + r (3 + r)}}}.\label{Ui3N}
\eea
The elements $U^N_{13,23,33}$ as functions of $r$ for $r \in (17.0, 19.0) $ are represented in Fig. \ref{Ui3N1d}.
For the case $r=18.0$ we get $U^N_{13}=0.141986, U^N_{23}=0.655677, U^N_{33}=0.741571$, as well as the following relations:
\bea
U^N_{12}&=& \fr{(0.290299  + 3.7183  i) - (3.07522 - 2.66143 i) x}{(0.905751 - 0.412844  i+ x)\sqrt{\fr{44.7731 + 52.8965 x + 45.1876 x^2}{0.990826 + 1.8115 x + x^2}}} ,\crn
U^N_{22} &=& \fr{(1.25788 + 4.15466  i) - (2.37253 - 3.46349 i) x}{(0.905751 - 0.412844 i+ x)\sqrt{\fr{44.7731 + 52.8965 x + 45.1876 x^2}{0.990826 + 1.8115 x + x^2}}},\crn
U^N_{32}&=& \fr{(1.20382 + 3.25121  i) - (2.04551 - 2.61508  i) x}{(0.915112 - 0.382591 i+ x)\sqrt{\fr{44.7731 + 52.8965 x + 45.1876 x^2}{0.990826 + 1.8115 x + x^2}}} .  \label{Ui2N}
\eea
In Fig. \ref{Ui2N1d}, we have plotted the values of $U^N_{12,22,32}$ as functions of $x$ with $x \in (-1.0, -0.6) $.

Now, taking the best fit experimental data on neutrino mass square difference, $\De m_{21}^2 =7.56 \times 10^{-5} \textrm{eV}^2$ and $\De m_{31}^2=2.55\times
10^{-3} \textrm{eV}^2$, given in  Ref. \protect\cite{deSalas:2017kay},
we obtain a solution\footnote{The system of equations has four physical solutions, however, they have no effect on the neutrino oscillation experiments.
So, here we only consider in
detail the case in Eq. (\ref{xzN}).}:
\be
x= -0.648025 , \hspace{0.5cm}z =4.74323\times 10^{-5}\, \textrm{eV},  \label{xzN}
\ee
and
\be
|m_2| =8.69482\times 10^{-3} \textrm{eV} , |m_3| =5.04976\times 10^{-2}\, \textrm{eV}.  \label{m23N}
\ee
The lepton mixing matrix in Eq.(\ref{RnuN}) then takes the form
\be
U^N=\left(
\begin{array}{ccc}
-0.804496 + 0.112566 i  & -0.0888079 + 0.551228 i    & 0.0429831 - 0.135324  i\\
0.084411 + 0.357999  i  &-0.0258114 + 0.623133  i    & 0.466887 + 0.460358 i\\
0.220085 + 0.39546  i    & 0.00350582 + 0.547072 i    & 0.709253 - 0.216534 i
\end{array}\right), \label{UN}
\ee
or
\be
|U^N|=\left(
\begin{array}{ccc}
0.812333 & 0.558336&0.141986\\
0.367816 & 0.623667 &0.655677\\
0.452577 & 0.547084 & 0.741571
\end{array}\right), \label{absUN}
\ee
which is consistent with the constraint on the absolute values of the
entries of the lepton mixing matrix given in Ref. \protect\cite{Gonzalez:2016}. The value of the
Jarlskog invariant determining the magnitude of CP violation in neutrino
oscillations in the model is then $J=2.69528\times 10^{-2}$. The obtained values for the charged lepton masses and leptonic mixing
parameters for the case of normal neutrino mass hierarchy are obtained starting from the following benchmark point:
\[
\begin{array}{c}
a_1^{(l) }\simeq 1.89, \hspace{0.5cm} a_2^{(l)
}\simeq 1.02, \hspace{0.5cm} a_3 ^{(l) }\simeq 0.88 , \hspace{0.5cm}
\al \simeq 51.57^{\circ }, \hspace{0.5cm} \bet \simeq 157.56^{\circ }\,.%
\end{array}%
\]
In what follows, we turn to the determination of the effective Majorana neutrino
mass parameter, which is proportional to the amplitude of neutrinoless
double beta ($0\nu \bet \bet $) decay. The effective Majorana neutrino
mass parameter reads $m_{ee}=\left\vert \sum_{k}U_{ek}^2 m_{\nu _{k}}\right\vert =3.6963$ meV, which is well below its current most
 strict experimentally upper limit $m_{ee}\leq 160$ meV, as follows from the constraint $T_{1/2}^{0\nu\bet \bet }(^{136}\mathrm{Xe})
 \geq 1.1\times 10^{26}$ yr at 90\% C.L obtained by the KamLAND-Zen experiment \cite{KamLAND-Zen:2016pfg}.
 Hence, our obtained effective Majorana neutrino mass parameter is beyond the reach of the present and forthcoming $0\nu \beta \beta $-decay experiments.

\section{Conclusions}
\label{conclusions}
We have built a viable theory based on the $SU(3)_C\times SU(3)_L\times U(1)_X$ gauge group, which is supplemented
by the $U(1)_{L_g}$ global lepton number symmetry and the $\De (27) \times Z_3\times Z_{16}$ discrete group, capable
 of providing a very good description of the low energy fermion flavor data. In our model, the spontaneous
  breakdown of the $\De (27) \times Z_3\times Z_{16}$ discrete symmetry takes place at very large energies,
  thus producing the observed SM fermion masses and mixings. The active neutrinos acquire small masses
  produced by the inverse seesaw mechanism mediated by three very light Majorana neutrinos. The lightness
   of the right handed Majorana neutrinos mediating the inverse seesaw mechanism is attributed to the fact that they obtain
    small masses from thirteen dimensional Yukawa terms involving a scalar singlet that acquires a vacuum expectation value
     at a scale much lower than the scale of breaking of the $\De (27) \times Z_3\times Z_{16}$ discrete group.
     In this model, small masses for active neutrinos are naturally generated since these masses are inversely proportional
     to powers of the large model cutoff $\La $ and feature a quadratic scaling on the very VEVs of the $SU(3)_L $
      singlet and $\De(27)$ triplet scalar fields $\zeta$ and $\Phi$. We perform a detailed analysis in the lepton sector, where
      the model is only viable for normal neutrino mass ordering, obtaining leptonic mixing parameters in excellent
       agreement with the experimental data and predicting $m_{ee}\simeq 3.7$ meV and $J\simeq 2.7\times 10^{-2}$.

\section*{Acknowledgments}

This research has been financially supported by Fondecyt (Chile), Grants
No.~1170803, CONICYT PIA/Basal FB0821, the Vietnam National Foundation for
Science and Technology Development (NAFOSTED) under grant number 103.01-2017.341. A.E.C.H is highly thankful to
the Institute of Physics, Vietnam Academy of Science and Technology for the friendly hospitality.

\end{document}